\newcommand{\CLASSINPUTtoptextmargin}{1.7cm}
\newcommand{\CLASSINPUTbottomtextmargin}{1.7cm}

\documentclass[conference]{IEEEtran}
\IEEEoverridecommandlockouts
\usepackage{amsmath,amssymb,amsfonts}
\usepackage{algorithmic}
\usepackage{graphicx}
\usepackage{textcomp}
\usepackage{xcolor}
\usepackage{listings}
\usepackage{url} 
\usepackage[utf8]{inputenc} 
\usepackage{booktabs}
\usepackage{etoolbox}
\usepackage{svg}

\usepackage{tikz}

\usepackage[numbers]{natbib}
\usepackage[frozencache,cachedir=.]{minted}

\usetikzlibrary{shapes.geometric, arrows, positioning, fit}
\usepackage[htt]{hyphenat}
\usepackage[many,minted]{tcolorbox}
\usepackage{caption}  
\usepackage{amsthm}   
\usetikzlibrary{shapes.geometric, arrows, positioning, fit}

\usepackage[htt]{hyphenat}

\usepackage[many,minted]{tcolorbox}
\setminted[julia]{fontsize=\footnotesize,linenos=true}
\setminted[python]{fontsize=\small,linenos=true}
\setminted[matlab]{fontsize=\small,linenos=true}
\setminted[common-lisp]{fontsize=\small,linenos=true}
\setminted[r]{fontsize=\small,linenos=true}
\setminted[shell]{fontsize=\small,linenos=true}
\setminted[yaml]{fontsize=\small,linenos=true}
\setminted[latex]{fontsize=\small,linenos=true}

\NewTCBListing[auto counter,number within=section]{myjulia}{ O{} O{} m m}{%
    listing only,
    enhanced jigsaw,
    breakable,
    colframe=gray!20,
    opacityback=0,
    attach boxed title to bottom center={yshift=-2mm},
    minipage boxed title,
    boxed title style={blanker},
    title={{\captionof{listing}{#3\label{lst:#4}}}},
    minted language=julia,
    minted options={numbersep=2.5mm,firstnumber=1,#2},
    #1,
    overlay={%
        \begin{tcbclipinterior}
            \fill[gray!25] (frame.south west) rectangle ([xshift=4mm]frame.north west);
        \end{tcbclipinterior}
    },
}
\NewTCBInputListing[use counter from=myjulia]{\myjuliainput}{ O{} O{} m m m}{%
    listing only,
    enhanced jigsaw,
    breakable,
    top=0.0\baselineskip,
    bottom=0.0\baselineskip,
    colframe=gray!20,
    opacityback=0,
    attach boxed title to bottom center={yshift=-2mm},
    minipage boxed title,
    boxed title style={blanker},
    title={{\captionof{listing}{#4\label{lst:#5}}}},
    minted language=julia,
    listing file={#3},
    minted options={numbersep=2.5mm,firstnumber=1,#2},
    overlay={%
        \begin{tcbclipinterior}
            \fill[gray!25] (frame.south west) rectangle ([xshift=4mm]frame.north west);
        \end{tcbclipinterior}
    },
    #1,
}

\NewTCBListing[auto counter,number within=section]{myjuliainline}{ O{} O{} m m}{%
    listing only,
    enhanced jigsaw,
    breakable,
    top=0.0\baselineskip,
    bottom=0.0\baselineskip,
    colframe=gray!20,
    opacityback=0,
    attach boxed title to bottom center={yshift=-2mm},
    minipage boxed title,
    boxed title style={blanker},
    title={{\captionof{listing}{#3\label{lst:#4}}}},
    minted language=julia,
    minted options={numbersep=2.5mm,firstnumber=1,#2},
    #1,
    overlay={%
        \begin{tcbclipinterior}
            \fill[gray!25] (frame.south west) rectangle ([xshift=4mm]frame.north west);
        \end{tcbclipinterior}
    },
}

\usepackage{threeparttable} 

\usepackage{pgfplots}
\pgfplotsset{compat=1.17}
\usepgfplotslibrary{polar}

\usepackage{array}

\newcommand*{\Data}{\mathcal{D}}

\DeclareUnicodeCharacter{03BD}{$\nu$}
\DeclareUnicodeCharacter{03BB}{$\lambda$}
\DeclareUnicodeCharacter{221A}{$\sqrt{}$}
\DeclareUnicodeCharacter{2218}{$\circ$}
\DeclareUnicodeCharacter{211D}{$\mathbb{R}$}
\DeclareUnicodeCharacter{00B2}{$^2$}
\DeclareUnicodeCharacter{2080}{$_0$}
\DeclareUnicodeCharacter{2081}{$_1$}
\DeclareUnicodeCharacter{1D40}{$^T$}
\DeclareUnicodeCharacter{03B1}{$\alpha$}
\DeclareUnicodeCharacter{03B2}{$\beta$}
\DeclareUnicodeCharacter{03BC}{$\mu$}
\DeclareUnicodeCharacter{03D5}{$\phi$}
\DeclareUnicodeCharacter{03C9}{$\omega$}
\DeclareUnicodeCharacter{03C4}{$\tau$}
\DeclareUnicodeCharacter{03BA}{$\kappa$}
\DeclareUnicodeCharacter{03B3}{$\gamma$}
\DeclareUnicodeCharacter{03F5}{$\epsilon$}
\DeclareUnicodeCharacter{207B}{$^{-}$}
\DeclareUnicodeCharacter{2248}{$\approx$}
\DeclareUnicodeCharacter{03B8}{$\theta$}
\DeclareUnicodeCharacter{03BC}{$\mu$}
\DeclareUnicodeCharacter{03C3}{$\sigma$}
\DeclareUnicodeCharacter{03C1}{$\rho$}
\DeclareUnicodeCharacter{2208}{$\in$}
\DeclareUnicodeCharacter{223C}{$\sim$}
\DeclareUnicodeCharacter{2264}{$\le$}
\DeclareUnicodeCharacter{2265}{$\ge$}
\DeclareUnicodeCharacter{221E}{$\infty$}
\DeclareUnicodeCharacter{27F9}{$\implies$}
\DeclareUnicodeCharacter{03C0}{$\pi$}
\DeclareUnicodeCharacter{1D62}{$_i$}
\DeclareUnicodeCharacter{209C}{$_t$}
\DeclareUnicodeCharacter{2081}{$_1$}
\DeclareUnicodeCharacter{2082}{$_2$}
\DeclareUnicodeCharacter{208A}{$_+$}
\DeclareUnicodeCharacter{2260}{$\ne$}
\DeclareUnicodeCharacter{1D4A9}{$\mathcal{N}$}
\DeclareUnicodeCharacter{2207}{$\nabla$}
\DeclareUnicodeCharacter{1D53C}{$\mathbb{E}$}
\DeclareUnicodeCharacter{03A3}{$\Sigma$}
\DeclareUnicodeCharacter{2713}{$\checkmark$}
\DeclareUnicodeCharacter{00D7}{$\times$}
\DeclareUnicodeCharacter{22EE}{$\vdots$}
\DeclareUnicodeCharacter{22EE}{$\vdots$}
\DeclareUnicodeCharacter{2145}{$\Data{}$}

\newtheoremstyle{usergroupstyle}
  {3pt}
  {3pt}
  {}
  {}
  {\bfseries}
  {:}
  {.5em}
  {}

\theoremstyle{usergroupstyle}

\newcommand{\usergroupref}[1]{\hyperref[#1]{UG~\ref*{#1}}}

\usepackage[capitalize,sort,compress]{cleveref}

\newcommand{\exref}[1]{\hyperref[#1]{Example~\ref*{#1}}}
\newcommand{\figref}[1]{\hyperref[#1]{Figure~\ref*{#1}}}
\newcommand{\secref}[1]{\hyperref[#1]{Section~\ref*{#1}}}
\newcommand{\lstref}[1]{\hyperref[#1]{Listing~\ref*{#1}}}
\newcommand{\tabref}[1]{\hyperref[#1]{Table~\ref*{#1}}}
\makeatletter
\patchcmd{\@maketitle}
  {\addvspace{0.5\baselineskip}\egroup}
  {\addvspace{-1\baselineskip}\egroup}
  {}
  {}
\makeatother

\def\BibTeX{{\rm B\kern-.05em{\sc i\kern-.025em b}\kern-.08em
    T\kern-.1667em\lower.7ex\hbox{E}\kern-.125emX}}

\begin{document}

\setcitestyle{square}
\hyphenation{PMCMC}

\title{State Space Model Programming in Turing.jl
}

\author{\IEEEauthorblockN{Tim Hargreaves\IEEEauthorrefmark{1},
Qing Li\IEEEauthorrefmark{1},
Charles Knipp\IEEEauthorrefmark{2}, 
Frederic Wantiez,
Simon J. Godsill\IEEEauthorrefmark{1},
Hong Ge\IEEEauthorrefmark{1}}
\IEEEauthorblockA{\IEEEauthorrefmark{1}Department of Engineering, University of Cambridge, UK \quad \IEEEauthorrefmark{2}Federal Reserve Board of Governors, USA\vspace{0.3em}}
}
\maketitle

\begin{abstract}
State space models (SSMs) are a powerful and widely-used class of probabilistic models for analysing time-series data across various fields, from econometrics to robotics. Despite their prevalence, existing software frameworks for SSMs often lack compositionality and scalability, hindering experimentation and making it difficult to leverage advanced inference techniques. This paper introduces \texttt{SSMProblems.jl} and \texttt{GeneralisedFilters.jl}, two Julia packages within the \texttt{Turing.jl} ecosystem, that address this challenge by providing a consistent, composable, and general framework for defining SSMs and performing inference on them. This unified interface allows researchers to easily define a wide range of SSMs and apply various inference algorithms, including Kalman filtering, particle filtering, and combinations thereof. By promoting code reuse and modularity, our packages reduce development time and improve the reliability of SSM implementations. We prioritise scalability through efficient memory management and GPU-acceleration, ensuring that our framework can handle large-scale inference tasks.

\end{abstract}

\section{Introduction}\label{intro}

State space models (SSMs) \cite{sarkka2023bayesian} are a fundamental framework for analysing sequential data across various fields, including finance (e.g., modelling stock prices) \cite{zeng2013state}, ecology (e.g., analysing population dynamics)  \cite{auger2021guide}, engineering (e.g., controlling robotic systems) \cite{friedland2005control}, and natural sciences  (e.g., predicting weather patterns) \cite{wendroth2014state}. An SSM models how a system changes over time by describing the evolution of an unobserved \textit{latent} Markov chain, $(x_t)_{t=0}^T$, and its relationship to (noisy) observations, $(y_t)_{t=1}^T$. In its most general form, an SSM is defined by three distributions that may depend on static model parameters $\theta$:
\begin{itemize}
    \item $p_\theta(x_0)$ — initialisation distribution
    \item $p_\theta(x_t | x_{t-1})$ — transition distribution
    \item $p_\theta(y_t | x_t)$ — observation distribution
\end{itemize}
for $\; t=1,\ldots,T$. Given these, the joint distribution of all $x_t$ and $y_t$ can be succinctly decomposed as,
\begin{equation}
p_\theta(x_{0:T}, y_{1:T}) = p_\theta(x_0) \prod_{t=1}^T p_\theta(x_t|x_{t-1}) p_\theta(y_t|x_t). \label{eq:ssm-decomp}
\end{equation}
This decomposition has two key characteristics. First, as $x_t$ is a Markov chain, the distribution of the latent dynamics' next state depends only on the current state. Secondly, after the latent state is conditioned, the observations are generated independently of each other.

A common objective when analysing SSMs is Bayesian filtering, which infers the current latent state given all observations up to the current time. This involves studying the posterior distribution $p_\theta(x_t|y_{1:t})$. A related task is Bayesian smoothing, which aims to infer the joint posterior of all latent states up to time $t$, $p_\theta(x_{0:t}|y_{1:t})$. While general Markov Chain Monte Carlo (MCMC) frameworks like Metropolis-Hastings or Hamiltonian Monte Carlo \cite{robert1999monte} (provided by probabilistic programming languages such as Stan \cite{carpenter2017stan} or PyMC \cite{abril2023pymc}) might seem applicable, they are often inefficient for SSMs. The sequential nature of the data and increasing dimensionality with time lead to a high computational burden and slow convergence \cite{betancourt2011geometry}. Instead, one should take advantage of the decomposition provided in Equation \ref{eq:ssm-decomp}, to generate the filtering (similarly, smoothing) distributions sequentially and recursively via a \emph{prediction step}:
$$p_\theta(x_t | y_{1:t-1}) = \int p_\theta(x_t | x_{t-1}) p_\theta(x_{t-1} | y_{1:t-1}) dx_{t-1},$$
and \emph{update step} (using Bayes' rule):
$$p_\theta(x_t | y_{1:t}) \propto p_\theta(y_t | x_t) p_\theta(x_t | y_{1:t-1}).$$
For some models, these steps can be performed in closed-form, but in a general setting, sequential Monte Carlo (SMC) methods, such as the particle filter \cite{godsill2019particle}, are typically employed.

In addition, there may be settings where the model parameters $\theta$ are unknown and must be inferred. This amounts to studying the joint posterior $p(x_{1:T}, \theta | y_{1:T})$, a task that can be accomplished using particle Markov chain Monte Carlo (PMCMC) methods \cite{andrieu2010particle} or SMC$^2$ \cite{chopin2013smc2}.

This work introduces two software packages, \texttt{SSMProblems.jl} and \texttt{GeneralisedFilters.jl}, that provide a unified interface for defining various SSMs and their inference algorithms, respectively. These packages are part of the \texttt{Turing.jl} \cite{ge2018turing} ecosystem, but offer a specific focus on optimised inference for SSMs, including efficient implementations of Kalman filtering and SMC methods. This contrasts with the more general Bayesian modelling approach of the original \texttt{Turing.jl} package. Section \ref{design} discusses our key design decisions and the advanced features our work supports compared to existing frameworks, before briefly introducing three case studies we include in the appendices, demonstrating our claims and providing practical use-cases of our framework.

\section{State Space Model Programming} \label{design}

\subsection{\texttt{SSMProblems.jl}: a minimal interface for specifying state-space models } \label{sec:interface}

\texttt{SSMProblems.jl} provides a minimal yet powerful interface for specifying a wide range of state space models (SSMs). 
SSMs are defined in \texttt{SSMProblems.jl} as a composition of two types representing the latent dynamics and observation process of the model. Such types may implement a \texttt{distribution} method for the initialisation, transition, and observation distributions defined in Section \ref{intro}, returning a Distributions.jl object. Alternatively, one may define \texttt{simulate} and \texttt{logdensity} methods for simulating from and computing log-densities of these distributions directly (with default implementations using \texttt{distribution}).

As a simple example, non-linear Gaussian dynamics could be defined as follows.
\myjuliainput[width=0.5\textwidth][firstline=1,lastline=8,xleftmargin=-0.9mm]{code/nonlinearGaussiandynamics.jl}{Definition of non-linear Gaussian dynamics}{nonlinear Gaussian dynamics}
\vspace{-0.5em}



Here, \texttt{kwargs...} is Julia specific nomenclature that we use as a convenient way of forwarding inference-time inputs, such as control or time variables, that enable time-inhomogeneous and controlled dynamics.

Our interface is designed to be lightweight, providing only the minimal functionality required to define SSMs. Despite its minimal design, our interface remains general-purpose, with additional model properties implemented by extending this foundational form. For example, linear Gaussian latent dynamics have a transition distribution of the form
$$p(x_t | x_{t-1}) \sim \mathcal{N}(A_t x_t + b_t, Q_t).$$
It is vital that the matrices/vectors $A_t, b_t, Q_t$ can be accessed directly so that closed-form inference using the Kalman filter can be performed. \texttt{SSMProblems.jl} makes these available by defining an abstract type, \texttt{LinearGaussianLatentDynamics}, that expects methods for \texttt{calc\_A}, \texttt{calc\_b} and \texttt{calc\_Q} to be implemented, with signatures similar to that of \texttt{distribution}. These methods can then be used directly by the Kalman filter algorithm for exact inference, or as ingredients for particle filters. In this sense, our state-space modelling interface is unified—one can define a linear Gaussian SSM once and expect both efficient closed-form and generic (e.g. particle filter) inference algorithms to work without additional implementation effort. This is desirable when comparing multiple candidate models, as the most general form of particle filtering can be applied to all of them, providing a consistent benchmark.

By splitting the SSM definition into latent dynamics and observation process components, the interface is modular, allowing the user to easily replace either component without redefining the entire model. This also allows for constructing complex hierarchical models that admit Rao-Blackwellisation \cite{murphy2001rao} as detailed in Appendix \ref{app:blackbox}.

Our modelling interface is also robust in that it is designed to enforce the consistency of latent states and observations' numeric value types via Julia's parametric types. This design makes models compatible with automatic differentiation \cite{margossian2019review}. It also allows for preallocation of particle storage, given that all types are known before running inference algorithms. These design choices ensure that \texttt{SSMProblems.jl} provides a flexible, robust, and efficient foundation for defining and working with SSMs.

\subsection{\texttt{GeneralisedFilters.jl}: a general-purpose interface for filtering algorithms} \label{sec:filtering_interface}

\texttt{GeneralisedFilters.jl} provides efficient implementations of various filtering algorithms using the standardised interface provided by \texttt{SSMProblems.jl}. Specifically, an inference algorithm is defined by writing \texttt{predict} and \texttt{update} methods that implement the recursive filtering steps defined in Section \ref{intro}. Importantly, these methods are defined in a unified way, each accepting and returning a distribution(-like) object, such as a Gaussian distribution or collection of weighted particles, representing the estimated predicted/filtered distribution. This eliminates the interface differences between, e.g., Kalman and particle filtering algorithms, simplifying the process of switching between different filtering techniques and allowing users to easily explore and compare their performance without significant code modifications. The update step also returns the incremental log-likelihood $\log p_\theta(y_t | y_{1:t-1})$ which can be used to compute the marginal log-likelihood $\log p_\theta(y_{1:t})$.

The modular design of these methods, which act as interchangeable generators of filtering distributions and likelihoods, allows for elegant composition of inference algorithms. A key example is that of Rao-Blackwellisation \cite{murphy2001rao}, where a conditional sub-structure of the SSM can be analytically marginalised (e.g. using the Kalman Filter) to reduce the variance of estimates compared to naive particle filtering. Our design allows us to easily implement a Rao-Blackwellised particle filter that can use any analytical model and corresponding closed-form inference algorithm for marginalisation, rather than only the standard choice of linear-Gaussian, as explored in Appendix \ref{app:blackbox}.

\begin{table}[ht]
\centering
\scriptsize
\begin{tabular}{@{}l@{\hspace{1.0em}}c@{\hspace{1.0em}}c@{\hspace{1.0em}}c@{\hspace{1.0em}}c@{}}
\toprule
\textbf{Feature} & \textbf{Ours} & \textbf{Birch} \cite{murray2018automated} & \textbf{StoneSoup} \cite{thomas2017open} & \textbf{particles} \cite{chopin2020introduction} \\
\midrule
Unified Interface & \checkmark & $\times$ & $\checkmark$ & $\times$  \\
Rao-Blackwellisation & \checkmark & \checkmark & $\times$ & $\times$   \\
PMCMC & \checkmark & $\times$ & $\times$ & $\checkmark$  \\
GPU-Acceleration & \checkmark & \checkmark & $\times$ & $\times$ \\
Automatic Differentiation & \checkmark & \checkmark & $\times$ & $\times$  \\
Sparse Particle Storage & \checkmark & $\times$ & $\times$ & $\times$  \\
\bottomrule
\\[-8.0pt]
\end{tabular}
\caption{Advanced feature comparison for SSM frameworks}
\label{tab:comparison}

\end{table}
\vspace{0.1cm}

Our inference algorithms are designed to scale to high-dimensional and streaming contexts, or be used in computationally intensive inference procedures such as particle MCMC. This was achieved by implementing GPU-accelerated versions of all algorithms and efficient methods for particle genealogy storage, such as Jacob et al.'s sparse path storage \cite{jacob2015path}. The performance gains from our GPU implementation can be found in Table \ref{tab:gpu} with experimental details provided in Appendix \ref{app:scalability}.

Table \ref{tab:comparison} compares and relates our work to three existing popular SSM frameworks.

\begin{table}[ht]
\centering
\scriptsize
\begin{tabular}{@{}l@{\hspace{1.0em}}c@{\hspace{1.0em}}c@{\hspace{1.0em}}c@{\hspace{1.0em}}c@{}}
\toprule
\textbf{Hardware} & \textbf{Mean Wall Time Per Step ($\pm$ 1 std. dev)} & \textbf{Relative Speed-Up} \\
\midrule
CPU & 678 ms $\pm$ 17 ms &  1.0 \\
GPU & 8.5 ms $\pm$ 3.1 ms& 79.4 &   \\
\bottomrule
\\[-8.0pt]
\end{tabular}
\caption{Runtime comparisons of CPU and GPU implementations of the Rao-Blackwellised particle filter ($N=10^5$)}
\label{tab:gpu}

\end{table}
\vspace{0.1cm}

Appendices C–E demonstrate the versatility of our state-space model programming framework through real-world applications, including multi-object tracking, trend inflation analysis, and data assimilation. These diverse case studies demonstrate the framework's ability to handle complex scenarios and provide robust solutions.

\clearpage
\onecolumn
\bibliographystyle{IEEEtran}
\bibliography{bibliography}

\begin{thebibliography}{10}
\providecommand{\url}[1]{#1}
\csname url@samestyle\endcsname
\providecommand{\newblock}{\relax}
\providecommand{\bibinfo}[2]{#2}
\providecommand{\BIBentrySTDinterwordspacing}{\spaceskip=0pt\relax}
\providecommand{\BIBentryALTinterwordstretchfactor}{4}
\providecommand{\BIBentryALTinterwordspacing}{\spaceskip=\fontdimen2\font plus
\BIBentryALTinterwordstretchfactor\fontdimen3\font minus \fontdimen4\font\relax}
\providecommand{\BIBforeignlanguage}[2]{{%
\expandafter\ifx\csname l@#1\endcsname\relax
\typeout{** WARNING: IEEEtran.bst: No hyphenation pattern has been}%
\typeout{** loaded for the language `#1'. Using the pattern for}%
\typeout{** the default language instead.}%
\else
\language=\csname l@#1\endcsname
\fi
#2}}
\providecommand{\BIBdecl}{\relax}
\BIBdecl

\bibitem{sarkka2023bayesian}
S.~S{\"a}rkk{\"a} and L.~Svensson, \emph{Bayesian filtering and smoothing}.\hskip 1em plus 0.5em minus 0.4em\relax Cambridge university press, 2023, vol.~17.

\bibitem{zeng2013state}
Y.~Zeng and S.~Wu, \emph{State-space models: Applications in economics and finance}.\hskip 1em plus 0.5em minus 0.4em\relax Springer, 2013, vol.~1.

\bibitem{auger2021guide}
M.~Auger-M{\'e}th{\'e}, K.~Newman, D.~Cole, F.~Empacher, R.~Gryba, A.~A. King, V.~Leos-Barajas, J.~Mills~Flemming, A.~Nielsen, G.~Petris \emph{et~al.}, ``A guide to state--space modeling of ecological time series,'' \emph{Ecological Monographs}, vol.~91, no.~4, p. e01470, 2021.

\bibitem{friedland2005control}
B.~Friedland, \emph{Control system design: an introduction to state-space methods}.\hskip 1em plus 0.5em minus 0.4em\relax Courier Corporation, 2005.

\bibitem{wendroth2014state}
O.~Wendroth, Y.~Yang, and L.~C. Timm, ``State-space analysis in soil physics,'' \emph{Application of Soil Physics in Environmental Analyses: Measuring, Modelling and Data Integration}, pp. 53--74, 2014.

\bibitem{robert1999monte}
C.~Robert, ``{Monte Carlo} statistical methods,'' 1999.

\bibitem{carpenter2017stan}
B.~Carpenter, A.~Gelman, M.~D. Hoffman, D.~Lee, B.~Goodrich, M.~Betancourt, M.~Brubaker, J.~Guo, P.~Li, and A.~Riddell, ``Stan: A probabilistic programming language,'' \emph{Journal of statistical software}, vol.~76, pp. 1--32, 2017.

\bibitem{abril2023pymc}
O.~Abril-Pla, V.~Andreani, C.~Carroll, L.~Dong, C.~J. Fonnesbeck, M.~Kochurov, R.~Kumar, J.~Lao, C.~C. Luhmann, O.~A. Martin \emph{et~al.}, ``Pymc: a modern, and comprehensive probabilistic programming framework in python,'' \emph{PeerJ Computer Science}, vol.~9, p. e1516, 2023.

\bibitem{betancourt2011geometry}
M.~Betancourt and L.~C. Stein, ``The geometry of hamiltonian {Monte Carlo},'' \emph{arXiv preprint arXiv:1112.4118}, 2011.

\bibitem{godsill2019particle}
S.~Godsill, ``Particle filtering: the first 25 years and beyond,'' in \emph{ICASSP 2019-2019 IEEE International Conference on Acoustics, Speech and Signal Processing (ICASSP)}.\hskip 1em plus 0.5em minus 0.4em\relax IEEE, 2019, pp. 7760--7764.

\bibitem{andrieu2010particle}
C.~Andrieu, A.~Doucet, and R.~Holenstein, ``Particle {M}arkov chain {Monte Carlo} methods,'' \emph{Journal of the Royal Statistical Society Series B: Statistical Methodology}, vol.~72, no.~3, pp. 269--342, 2010.

\bibitem{chopin2013smc2}
N.~Chopin, P.~E. Jacob, and O.~Papaspiliopoulos, ``Smc2: an efficient algorithm for sequential analysis of state space models,'' \emph{Journal of the Royal Statistical Society Series B: Statistical Methodology}, vol.~75, no.~3, pp. 397--426, 2013.

\bibitem{ge2018turing}
H.~Ge, K.~Xu, and Z.~Ghahramani, ``Turing: a language for flexible probabilistic inference,'' in \emph{International conference on artificial intelligence and statistics}.\hskip 1em plus 0.5em minus 0.4em\relax PMLR, 2018, pp. 1682--1690.

\bibitem{murphy2001rao}
K.~Murphy and S.~Russell, ``Rao-{Black}wellised particle filtering for dynamic {Bayesian} networks,'' in \emph{Sequential Monte Carlo methods in practice}.\hskip 1em plus 0.5em minus 0.4em\relax Springer, 2001, pp. 499--515.

\bibitem{margossian2019review}
C.~C. Margossian, ``A review of automatic differentiation and its efficient implementation,'' \emph{Wiley interdisciplinary reviews: data mining and knowledge discovery}, vol.~9, no.~4, p. e1305, 2019.

\bibitem{murray2018automated}
L.~M. Murray and T.~B. Sch{\"o}n, ``Automated learning with a probabilistic programming language: Birch,'' \emph{Annual Reviews in Control}, vol.~46, pp. 29--43, 2018.

\bibitem{thomas2017open}
P.~A. Thomas, J.~Barr, B.~Balaji, and K.~White, ``An open source framework for tracking and state estimation ('stone soup'),'' in \emph{Signal Processing, Sensor/Information Fusion, and Target Recognition XXVI}, vol. 10200.\hskip 1em plus 0.5em minus 0.4em\relax SPIE, 2017, pp. 62--71.

\bibitem{chopin2020introduction}
N.~Chopin, O.~Papaspiliopoulos \emph{et~al.}, \emph{An introduction to sequential {Monte Carlo}}.\hskip 1em plus 0.5em minus 0.4em\relax Springer, 2020, vol.~4.

\bibitem{jacob2015path}
P.~E. Jacob, L.~M. Murray, and S.~Rubenthaler, ``Path storage in the particle filter,'' \emph{Statistics and Computing}, vol.~25, pp. 487--496, 2015.

\bibitem{andrieu2009pseudo}
\BIBentryALTinterwordspacing
C.~Andrieu and G.~O. Roberts, ``The pseudo-marginal approach for efficient monte carlo computations,'' \emph{The Annals of Statistics}, vol.~37, no.~2, pp. 697--725, 2009. [Online]. Available: \url{http://www.jstor.org/stable/30243645}
\BIBentrySTDinterwordspacing

\bibitem{cappeInferenceHiddenMarkov2006}
O.~Capp{\'e}, E.~Moulines, and T.~Ryden, \emph{Inference in {{Hidden Markov Models}}}.\hskip 1em plus 0.5em minus 0.4em\relax Springer Science \& Business Media, Apr. 2006.

\bibitem{terejanu2008extended}
G.~A. Terejanu \emph{et~al.}, ``Extended {Kalman} filter tutorial,'' \emph{University at Buffalo}, vol.~27, 2008.

\bibitem{besard2016high}
T.~Besard, P.~Verstraete, and B.~De~Sutter, ``High-level gpu programming in julia,'' \emph{arXiv preprint arXiv:1604.03410}, 2016.

\bibitem{leal2015motchallenge}
L.~Leal-Taix{\'e}, ``Motchallenge 2015: Towards a benchmark for multi-target tracking,'' \emph{arXiv preprint arXiv:1504.01942}, 2015.

\bibitem{li2023adaptive}
Q.~Li, R.~Gan, J.~Liang, and S.~J. Godsill, ``An adaptive and scalable multi-object tracker based on the non-homogeneous {P}oisson process,'' \emph{IEEE Transactions on Signal Processing}, vol.~71, pp. 105--120, 2023.

\bibitem{gan2024variational}
R.~Gan, Q.~Li, and S.~J. Godsill, ``Variational tracking and redetection for closely-spaced objects in heavy clutter,'' \emph{IEEE Transactions on Aerospace and Electronic Systems}, 2024.

\bibitem{li2023scalable}
Q.~Li, R.~Gan, and S.~Godsill, ``A scalable {Rao-Blackwellised} sequential {MCMC} sampler for joint detection and tracking in clutter,'' in \emph{2023 26th International Conference on Information Fusion (FUSION)}.\hskip 1em plus 0.5em minus 0.4em\relax IEEE, 2023, pp. 1--8.

\bibitem{stock2007}
J.~H. Stock and M.~W. Watson, ``Why has us inflation become harder to forecast?'' \emph{Journal of Money, Credit and banking}, vol.~39, pp. 3--33, 2007.

\bibitem{stock2016}
------, ``Core inflation and trend inflation,'' \emph{Review of Economics and Statistics}, vol.~98, no.~4, pp. 770--784, 2016.

\bibitem{carrassi2018data}
A.~Carrassi, M.~Bocquet, L.~Bertino, and G.~Evensen, ``Data assimilation in the geosciences: An overview of methods, issues, and perspectives,'' \emph{Wiley Interdisciplinary Reviews: Climate Change}, vol.~9, no.~5, p. e535, 2018.

\bibitem{lorenz1963deterministic}
E.~N. Lorenz, ``Deterministic nonperiodic flow,'' \emph{Journal of atmospheric sciences}, vol.~20, no.~2, pp. 130--141, 1963.

\bibitem{rackauckas2017differentialequations}
C.~Rackauckas and Q.~Nie, ``Differentialequations. jl--a performant and feature-rich ecosystem for solving differential equations in julia,'' \emph{Journal of open research software}, vol.~5, no.~1, pp. 15--15, 2017.

\end{thebibliography}

\onecolumn
\appendices

\section{Supplementary materials for Rao-Blackwellisation and PMCMC} \label{app:blackbox}

In this appendix, we elaborate on the unified filtering interface provided by \texttt{GeneralisedFilters.jl} and demonstrate how this leads to simple implementations of particle MCMC algorithms \cite{andrieu2010particle} and Rao-Blackwellised particle filters \cite{murphy2001rao}.

\subsection{Inference Interface Design}

As noted in the Section \ref{sec:filtering_interface}, the \texttt{predict} and \texttt{update} functions specified by the interface are designed in a modular fashion, each accepting and returning a distribution-like object. In the context of the particle filter, this is a weighted collection of particles, represented in code as a \texttt{ParticleState} and defined in Listing 2.

The corresponding \texttt{update} method for the particle filter is also shown. This acts upon a \texttt{ParticleContainer}, a pre-allocated unit of memory used for storing the proposed and filtered distributions for the current time step, generated by \texttt{predict} and \texttt{update} respectively. Note, that the \texttt{update} method additionally returns the incremental log-likelihood $p_\theta(y_t|y_{1:t-1})$.

\begin{myjuliainline}[][firstline=1,xleftmargin=0.0mm]{Definitions of the distribution-like object and update method for the bootstrap particle filter}{update}
mutable struct ParticleState{PT,WT<:Real}
    particles::Vector{PT}
    log_weights::Vector{WT}
end

mutable struct ParticleContainer{T,WT}
    filtered::ParticleState{T,WT}
    proposed::ParticleState{T,WT}
    ancestors::Vector{Int}
end

function update(
    model::StateSpaceModel{T},
    filter::BootstrapFilter,
    step::Integer,
    states::ParticleContainer,
    observation;
    kwargs...,
) where {T}
    log_increments = map(
        x -> SSMProblems.logdensity(model.obs, step, x, observation; kwargs...),
        collect(states.proposed),
    )

    states.filtered.log_weights = states.proposed.log_weights + log_increments
    states.filtered.particles = states.proposed.particles

    return states, logmarginal(states)
end
\end{myjuliainline}

This pattern is replicated across all filtering algorithms that implement our interface. A consequence of this design choice is that inference algorithms can be treated as ``black-box" generators of filtering distributions and corresponding likelihoods. This encapsulation allows arbitrary inference algorithms to be embedded within other functions and algorithms.

As a simple, elucidating example, our interface implements a \texttt{step} function, used to perform a combined predict-update step of a filtering algorithm. Due to the modular form of \texttt{predict} and \texttt{update}, this code—shown in Listing 3—can be written generically and shared between all filtering algorithms.

\begin{myjuliainline}[][firstline=1,xleftmargin=0.0mm]{Definition of \texttt{GeneralisedFilters.jl}'s generic \texttt{step} function}{step}
function step(
    rng::AbstractRNG,
    model::AbstractStateSpaceModel,
    alg::AbstractFilter,
    iter::Integer,
    state,
    observation;
    kwargs...,
)
    proposed_state = predict(rng, model, alg, iter, state; kwargs...)
    filtered_state, ll = update(model, alg, iter, proposed_state, observation; kwargs...)

    return filtered_state, ll
end
\end{myjuliainline}

\subsection{Particle MCMC Methods}

Of more practical interest, this design results in a simple way of implementing particle MCMC algorithms such as particle-marginal Metropolis-Hastings (PMMH) \cite{andrieu2010particle}. As with the generic \texttt{step} function, \texttt{GeneralisedFilters.jl} provides a generic \texttt{filter} function that computes the final filtering distribution and corresponding log-evidence, $p_\theta(y_{1:T})$ for an SSM, given a sequence of observations. In the context of particle filtering methods, this will be an unbiased estimate of the true log-evidence, which can be used to compute the acceptance ratio in a pseudo-marginal Metropolis-Hastings \cite{andrieu2009pseudo} implementation.

In fact, the log-evidence estimates from \texttt{GeneralisedFilters.jl} can be directly used within a \texttt{Turing.jl} model, taking advantage of the performant and well-tested implementation of Metropolis-Hastings it provides. A simplified example of how this could be used to perform inference on the noise variance parameters of the non-linear Gaussian model introduced in Section \ref{sec:interface} using a bootstrap particle filter is shown in Listing 4.

\begin{myjuliainline}[][firstline=1,xleftmargin=0.0mm]{Using \texttt{GeneralisedFilters.jl} within a \texttt{Turing.jl} model}{turing}
@model function pmmh_model(N, ys)
    # Priors
    σx ~ InverseGamma(2.0, 3.0)
    σy ~ InverseGamma(2.0, 3.0)

    f(x) = 0.9x  # arbitrary choice of dynamics
    model = StateSpaceModel(
        NLGDynamics(f, σx),
        LinearGaussianObservation(σy),
    )

    _, ll = filter(model, BootstrapFilter(N), ys)
    Turing.@addlogprob! ll

    return nothing
end
\end{myjuliainline}

Importantly, due to the modularity of the filtering interface, the same code for PMMH can be used regardless of the specific type of particle filter. That is, by simply changing \texttt{BootstrapFilter} to another type of particle filter, such as a guided, auxiliary, or even Rao-Blackwellised particle filter, one can run PMMH using a new inner algorithm with no other code modifications required. Further, if the model in question also had a linear Gaussian form, replacing the \texttt{BootstrapFilter} with a \texttt{KalmanFilter} would result in an exact (as opposed to pseudo-marginal) Metropolis-Hastings implementation.

\subsection{Rao-Blackwellisation}

The modular structure of our interface can also be utilised to easily implement a general version of the Rao-Blackwellised particle filter (RBPF) \cite{murphy2001rao}. Before we introduce this, we must explain how \texttt{SSMProblems.jl} implements models with a hierarchical structure.

A state space model may be suitable for Rao-Blackwellisation if its latent states $x_t$ can be decomposed into two components $x_t = (u_t, z_t)$ such that $u_t$ is an independent Markov chain and the observations only depend on $z_t$. In other words, we can decompose the joint density as,
\begin{equation}
p_\theta(x_{0:T}, y_{1:T}) = p_\theta(x_0) \prod_{t=1}^T p_\theta(u_t|u_{t-1})p_\theta(z_t|z_{t-1}, u_{t}) p_\theta(y_t|z_t).
\end{equation}

We call such state space models \textit{hierarchical} due to the fact that conditioning on fixed values of $(u_t)_{t=1}^T$ results in another, nested SSM. Such models are of particular interest when the inner model (after conditioning on $u_{1:T}$) has a special form (e.g., linear-Gaussian) that admits closed-form inference. In this case, one can utilise a Rao-Blackwellised particle filter, in which the ``outer"" states $u_t$ are inferred using a particle filter, with the ``inner`` states $z_t$ inferred using a closed-form inference algorithm, such as the Kalman filter, reducing the variance of generated estimates.

\texttt{SSMProblems.jl} represents hierarchical SSMs by combining two latent dynamics objects for $(u_t)_{t=1}^T$ and $(z_t)_{t=1}^T$, respectively, with an observation process. Dependencies of $(z_t)_{t=1}^T$ on the ``outer'' process $(u_t)_{t=1}^T$ can be made by using the keyword arguments \texttt{prev\_outer} and \texttt{new\_outer} when defining the dynamics, which at time $t$, represent the values $u_{t-1}$ and $u_t$ respectively. These keyword arguments are passed in by the filtering algorithm at inference time. Listing 5 shows an example of this process, defining the covariance matrix for a linear Gaussian ``inner'' dynamics that is scaled by a non-linear ``outer'' dynamics.

A hierarchical SSM is then defined by first combining the ``inner'' dynamics and observation process into a (conditional/nested) SSM, and then combining this with the ``outer'' dynamics using the ``HierarchicalSSM'' constructor, as also shown in Listing 5.

\begin{myjuliainline}[][firstline=1,xleftmargin=0.0mm]{Definition of conditional dynamics and a hierarchical state space model}{conditional_dynamics}
function GeneralisedFilters.calc_Q(dyn::InnerDynamics, ::Integer; new_outer, kwargs...)
    dyn.Q * new_outer
end

struct HierarchicalSSM{T<:Real,OD<:LatentDynamics,M<:AbstractStateSpaceModel} <: AbstractStateSpaceModel
    outer_dyn::OD
    inner_model::M
end
\end{myjuliainline}

Although many papers only consider Rao-Blackwellisation in the context of a linear Gaussian sub-model filtered using the Kalman filter, the technique is far more general, and we have designed our interface with this in mind. For example, one may have a SSM with a \emph{discrete} conditional sub-model (i.e. Hidden Markov Model) that can be filtered exactly using the forward algorithm \cite{cappeInferenceHiddenMarkov2006}. Going further, one could even use a Rao-Blackwellised particle filter when the ``inner'' model does not admit a closed-form filtering distribution but can be approximated as such, for example by using the extended Kalman filter \cite{terejanu2008extended}.

This generality is trivial to implement using the \texttt{GeneralisedFilters.jl} interface. Listing 6 first demonstrates how a Rao-Blackwellised particle filter is defined as an \texttt{RBPF} object. Importantly, this accepts an arbitrary filtering algorithm as a parameter, meaning that this single type can represent any generic RBPF algorithm. For this purpose, the \texttt{update} method for the RBPF (also shown in Listing 6) is completely independent of the specific inner filtering algorithm, leveraging Julia's powerful multiple dispatch, to call the specific \texttt{update} method of the inner filtering algorithm for marginalisation (line 12) without requiring knowledge of its details. The returned incremental log-likelihood is all that is required by the ``outer'' particle filter to update the particle weights, which is always provided inference algorithms subscribing to our interface.

\begin{myjuliainline}[][firstline=1,xleftmargin=0.0mm]{Definition and \texttt{update} method for the Rao-Blackwellised particle filter}{update method}
struct RBPF{F<:AbstractFilter,RS<:AbstractResampler} <: AbstractFilter
    inner_algo::F
    N::Int
    resampler::RS
end

function update(
    model::HierarchicalSSM{T}, algo::RBPF, t::Integer, states, obs; kwargs...
) where {T}
    log_increments = similar(states.filtered.log_weights)
    for i in 1:(algo.N)
        states.filtered.particles[i].u, log_increments[i] = update(
            model.inner_model,
            algo.inner_algo,
            t,
            states.proposed.particles[i].z,
            obs;
            new_outer=states.proposed.particles[i].u,
            kwargs...,
        )

        states.filtered.particles[i].u = states.proposed.particles[i].u
    end

    states.filtered.log_weights = states.proposed.log_weights + log_increments

    return states, logmarginal(states)
end
\end{myjuliainline}

Inference is performed using the exact same interface as the standard SSM, allowing for use in PMCMC algorithms as described above with no additional effort.

\newpage

\section{Supplementary materials for GPU-acceleration experiments} \label{app:scalability}

\texttt{GeneralisedFilters.jl} is notable for the inclusion of GPU-accelerated versions of numerous filtering algorithms, taking advantage of Julia's native support for CUDA programming via \texttt{CUDA.jl} \cite{besard2016high}. At the time of writing, the package has performant GPU-accelerated implementations of batched versions of the Kalman filter, square-root Kalman filter and forward algorithm, as well as the particle filter with a range of resampling algorithms. Further, due to the composability properties of our inference interface, it is trivial to be combine these algorithms together to form a GPU-accelerated Rao-Blackwellised particle filter (RBPF) \cite{murphy2001rao} using any of the aforementioned closed-form inference algorithms for marginalisation.

We will demonstrate the efficiency gains that come from GPU-acceleration by using the RBPF as a case study. This is representative inference algorithm as it features both the linear-algebraic operations of most closed form algorithms with particle filter resampling, that is less naturally suited to GPU-acceleration. The experiments were performed on a workstation PC with a 24-core AMD Ryzen Threadripper 7960X CPU and an Nvidia RTX 4090 GPU, both of which are comparable consumer-grade hardware.

To ensure the validity of our experiment, we use a fully linear-Gaussian model for which we can compute closed form ground truth filtering distributions using the Kalman filter, but treat a subset of the dimensions as if they have no special form. Specifically, writing $D_x, D_y$ for the dimensions of our latent states and observations, respectively, and splitting our latent state into $x_t = (u_t, z_t)$ with $u_t \in \mathbb{R}^{D_u}, z_t \in \mathbb{R}^{D_z}, D_u + D_z = D_x$, our model is given by,
\begin{align}
p(x_t | x_{t-1}) &= \mathcal{N}(Ax_{t-1}, Q), \\
p(y_t | x_t) &= \mathcal{N}(Hx_t, R),
\end{align}
where $A$, $H$, $Q$ have the block matrix forms,
\begin{align}
A = \begin{pmatrix}
A_{11} & 0_{D_u \times D_z} \\
A_{21} & A_{22}
\end{pmatrix}, \quad
H = \begin{pmatrix} 
0_{D_u \times D_y} & H_{2} \\
\end{pmatrix}, \quad
Q = \begin{pmatrix}
Q_{11} & 0_{D_u \times D_z} \\
0_{D_z \times D_u} & Q_{22}
\end{pmatrix}.
\end{align}

The structure of these matrices means that we can write our model in a hierarchical form,
\begin{align}
p(u_t | u_{t-1}) &= \mathcal{N}(A_{11}x_{t-1}, Q_{11}), \\
p(z_t | z_{t-1}, u_{t-1}) &= \mathcal{N}(A_{22}z_{t-1} + A_{21}u_{t-1}, Q_{22}) \\
p(y_t | z_t, u_t) = p(y_t | z_t) &= \mathcal{N}(H_{2}z_t , R)
\end{align}

For the sake of our experiment, we treat $u_t$ as if we do not know that it follows linear Gaussian dynamics, using a particle filter for these dimensions with the inference of $z_t$ marginalised out using a Kalman filter, resulting in a Rao-Blackwellised particle filter.

\begin{figure}[htp!]
    \centering
    \includegraphics[width=0.7\linewidth]{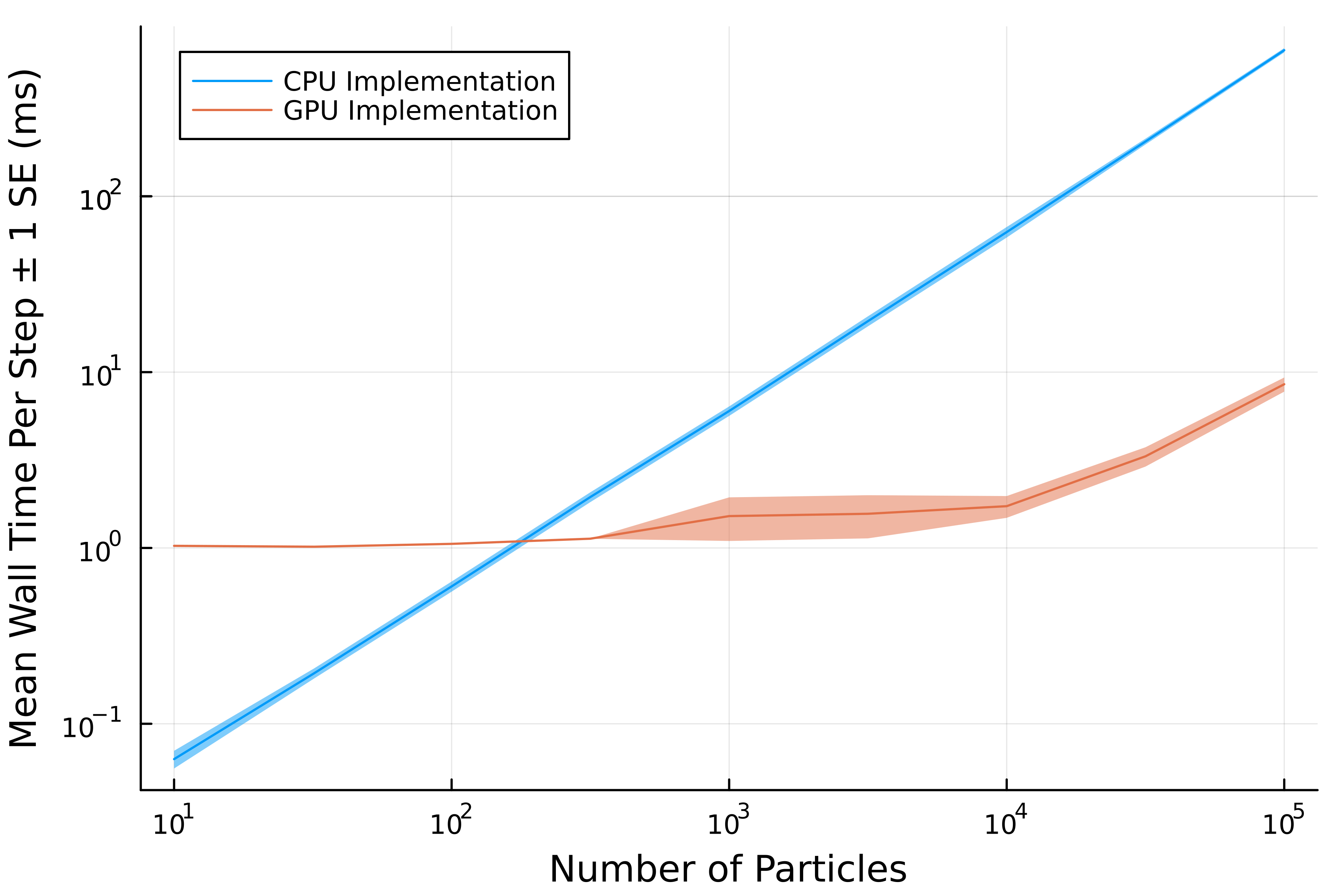}
    \caption{Benchmark results for CPU and GPU implementations of the Rao-Blackwellised particle filter}
    \label{fig:gpu_bench}
\end{figure}

We benchmark the CPU and GPU implementations of our RBPF over varying numbers of particles (denoted $N$) with $D_u = 2, D_z = 3, D_y = 2$ and randomly generated dense model matrices. The mean wall times per predict-update step are presented in Figure \ref{fig:gpu_bench}, on a log-log scale with standard error bars included.

As would be expected, for small particle counts ($N < 100$), the CPU implementation is the most performant, as there is not enough parallelism present to take full advantage of the GPU. However, whilst the wall time for the CPU implementation increases roughly linearly from this point, the runtime for the GPU implementation is almost constant due as addition work is computed in parallel. This is until the GPU becomes saturated at around $N=10^4$. At this point, the GPU implementation is roughly $40\times$ more performant than the CPU implementation, a gap that it maintained—and even slightly widened—as $N$ increases further. Exact figures for the case $N=10^5$ are given in Table \ref{tab:gpu}.

It is typically advised to use a minimum of 1000 particles when performing particle filtering, with more required as the dimension of the problem increases. This $40\times$ performance gain is therefore reflective of what a typical practitioner may expect.

Whilst we also benchmarked a multi-threaded CPU implementation of the RBPF with parallelism across particles, we found that this was no more performant than the serial CPU version. This is likely due to our serial implementation already taking advantage of multi-threaded BLAS operations.

\clearpage

\section{supplementary materials for multi object tracking case study}
Multi-object tracking (MOT) is a core problem in surveillance, aimed at estimating the trajectories of multiple objects over time from noisy sensor measurements. It is commonly formulated as a state-space model, where the hidden states represent the objects' positions and velocities, and the observations are noisy detections from sensors, often complicated by clutter and missed detections. A detailed formulation of the state-space model for tracking is provided in Section \ref{app:tracking}.

In this case study, we implement a multi-object tracker using \textit{SSMProblem.jl} and validate it with the MOTChallenge dataset in \cite{leal2015motchallenge}. Details of the experiment settings and implementation using \textit{SSMProblem.jl} are provided in Section \ref{sec:tracking SSMProblem}. As shown in Figure \ref{fig:tracking result}, the tracker successfully tracks all objects despite the data association uncertainty.

\subsection{State space model for multi object tracking}\label{app:tracking}
\subsubsection{Dynamic model}
In the MOT scenario, the dynamic model describes how the state of each object (such as position and velocity) evolves over time. Assume that each object's states evolving according to an independent linear Gaussian transition model: 
\begin{equation}\label{eq: dynamic transition}
    p(X_n|X_{n-1})
    =\mathcal{N}(X_{n};A_{n}X_{n-1},Q_{n}). 
\end{equation}
where $A_{n}$ the state transition matrix, $Q_{n}$ is Gaussian process noise covariance, and object state $X_{n}$ includes object's position and velocity.
\subsubsection{Measurement model}\label{sec:tracking measurement model}
In this case study, a point measurement model is used where each object generates one or zero measurements (due to possible missed detections). The measurements $Y_{n}=[Y_{n,1},...,Y_{n,M_n}]$ contain detections either from objects or clutter. At each time step, a target generates a measurement with probability $p_D\in(0,1]$ to generate a measurement, and the generated measurement follows the linear Gaussian model:
\begin{equation} 
\ell(Y_{n,j}|X_{n})=
    \mathcal{N}(Y_{n,j};H X_{n},R_{n})
\label{measurement model}
\end{equation} 
where $R_{n}$ represents the measurement noise covariance.

Clutter is modelled by a Poisson process with rate $\Lambda_0$. The total number of measurements $M_n$ follows a Poisson distribution with a rate $\Lambda_0$, and
the clutter measurement is uniformly distributed in the observation area of volume V:
\begin{equation} \label{eq:clutter}
p(M_n) = e^{-\Lambda_0} \frac{(\Lambda_0)^{M_n}}{M_n!}, \quad \ell_c(Y_{n,j})=
    {1}/{V}
\end{equation} 

\subsubsection{Data association}
Tracking multiple objects requires associating measurements with their corresponding objects. Assume the point measurement model in Section \ref{sec:tracking measurement model}. Let $\theta_{n}=[\theta_{n,1},...,\theta_{n, K}]$ represent the association vector, with each entry being
\begin{equation} \label{eq: asso define}
    \theta_{n,i} = \begin{cases} 
m \in \{1, \dots, M_n\}, & \text{if at time } k, \text{the $i$-th target generates a measurement } m  \\
0, & \text{if at time } k, \text{the $i$-thtarget does not generate a measurement}
\end{cases}
\end{equation}
Thus, $\theta_{n}$ represents a feasible association between
$M_n$ measurements and $K$ targets and we let $\Theta_{n}$ be a set of all feasible (joint) association events at time $n$. 
This association must satisfy two constraints: 1) An observation is associated with at most one target; 2) A target is associated with at most one observation.
For non-homogeneous Poisson process measurement model and its data association formulation, please refer to \cite{li2023adaptive}. 
\subsection{SSMProblems.jl for Tracking
}\label{sec:tracking SSMProblem}
The implementation of multi-object tracking using \textit{SSMProblem.jl} consists of several key modules. First, the \texttt{MultiTargetDynamics} is defined as a subtype of \texttt{LinearGaussianLatentDynamics} from the \textit{SSMProblem.jl} package, which models the state transitions of multiple objects’ dynamics as a linear Gaussian model, specifying state transition matrices, initial distributions, and process noise covariances. 
\myjuliainput[][firstline=1,lastline=9]{code/tracking.jl}{Multi-Target Dynamics}{tracking}

To model an observation process that consists of both target detections and clutter, we adopt a composition-based approach and define \texttt{ClutterAndMultiTargetObservations}, which represents a combined observation process involving both targets and clutter. Specifically, 
the \texttt{MultiTargetObservations}  handles the relationship between the hidden states and the sensor measurements, with a observation model introduced in \eqref{measurement model}. To address false alarms and clutter in the measurement data, the \texttt{ClutterObservations} module models clutter as a Poisson process as defined in \eqref{eq:clutter}. 
The \texttt{SSMProblems.simulate} function for \texttt{ClutterAndMultiTargetObservations} performs forward simulation by first simulating the target observations using the linear Gaussian model and then generating clutter following a homogenous Poisson process.
\myjuliainput[][firstline=11,lastline=34]{code/tracking.jl}{Observations with Clutter}{Observations with Clutter}




For data association, the \texttt{GlobalNearestNeighborAssociator} module applies the Hungarian algorithm to match predicted object states with observed measurements. This minimises the total association cost by calculating the distance between predicted states and measurements, ensuring that each target is correctly paired with its corresponding detection or labelled as undetected. The detailed implementation is given in the following codes. More sophisticated data association solution such as using variational inference \cite{gan2024variational} and Monte Carlo methods \cite{li2023scalable}, will be integrated into current scheme in future work.
\myjuliainput[][firstline=36,lastline=54]{code/tracking.jl}{Data association}{Data association}
    



Lastly the filtering is handled using a \texttt{Kalman Filter}, which recursively estimates the current states of objects based on the associated measurements. The filtering process operates within the \textit{SSMProblem.jl} framework, performing both prediction and update steps to continuously refine object trajectories.

\subsection{Settings and results for Motchallenge Dataset}
\begin{figure}[htp!]
    \centering
    \includegraphics[width=0.5\linewidth]{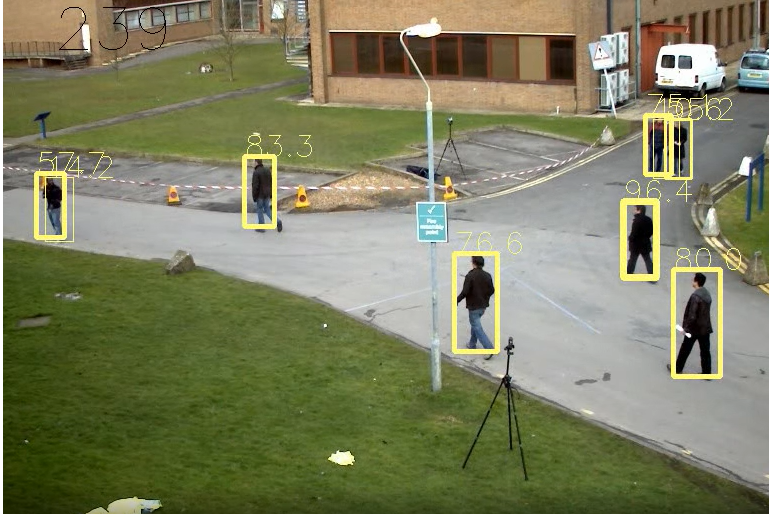}
    \caption{Original detections of 7 objects at one time step; yellow rectangles are detections including object detections and false detections}
    \label{fig:mot}
\end{figure}

\begin{figure}[htp!]
    \centering
    \includegraphics[width=0.5\linewidth]{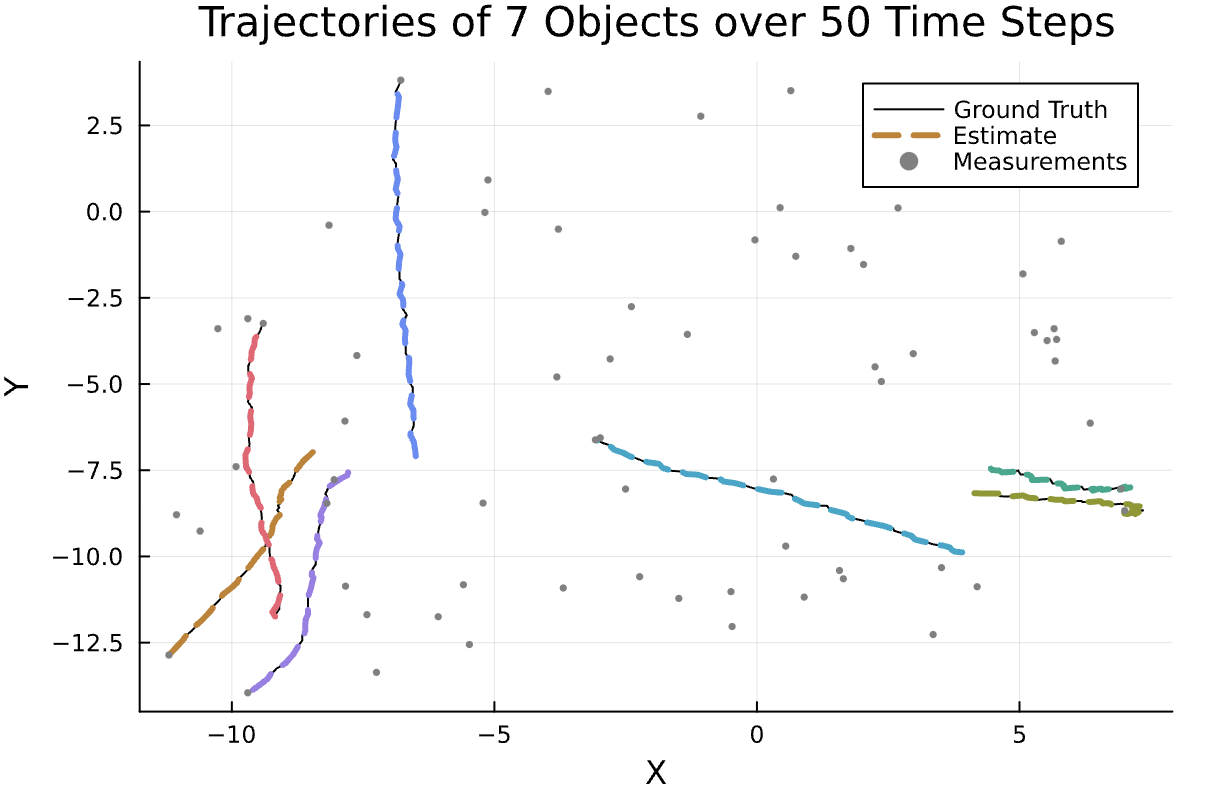}
    \caption{Ground-truth and estimated trajectories of 7 objects over 50 time steps; grey dots are total measurements at one time step including object detections and clutters}
    \label{fig:tracking result}
\end{figure}

In this case study, we use the MOTChallenge 2015 dataset, and the ground-truth video sequence can be found at \url{https://motchallenge.net/vis/PETS09-S2L1/det/}. Figure \ref{fig:mot} shows all detections at a single time step. To create a more challenging scenario and test the tracker’s data association capabilities, we introduce uniformly distributed clutter as described in Section \ref{sec:tracking measurement model} and an example of all detections at one time step can be seen in Figure \ref{fig:tracking result}. To perform the tracking task, we assume the underlying dynamical model for each object follows a near-constant velocity model. The system dynamics and measurement models are parameterised as follows:
\[
A = \begin{bmatrix}
1 & \tau & 0 & 0 \\
0 & 1 & 0 & 0 \\
0 & 0 & 1 & \tau \\
0 & 0 & 0 & 1
\end{bmatrix}, Q = \begin{bmatrix}
\frac{1}{3} & \frac{1}{2} & 0 & 0 \\
\frac{1}{2} & 1 & 0 & 0 \\
0 & 0 & \frac{1}{3} & \frac{1}{2} \\
0 & 0 & \frac{1}{2} & 1
\end{bmatrix}, H = \begin{bmatrix}
1 & 0 \\
0 & 0 \\
0 & 1 \\
0 & 0
\end{bmatrix}^\top, R = \begin{bmatrix}
0.3 & 0 \\
0 & 0.3
\end{bmatrix}
\]
where $A$ is the state transition matrix and $\tau$ is time interval, $Q$ is the process noise covariance matrix, $H$ is the observation matrix, and $R$ is the observation noise covariance matrix.

The estimated result is shown in Figure \ref{fig:tracking result}, where the black lines are ground truth trajectories, and colored dotted line are estimated trajectories. It shows that all objects are tracked closely to the true trajectories thus demonstrating the effectiveness of the tracker.

\newpage

\section{Supplementary materials for trend inflation case study}

\subsection{Unobserved components models}

\subsubsection{Local level trend model} One of the simplest trend dynamics is the weakly stationary random walk. In essence this is an order 1 autoregressive process with a single unit root. Let $x_{1:T}$ be a sequence of latent states (the underlying trend) measured by $y_{1:T}$.
\begin{align}
    y_{t} &= x_{t} + \eta_{t} \\
    x_{t+1} &= x_{t} + \varepsilon_{t}
\end{align}

\subsubsection{Stochastic volatility} In a linear Gaussian setting $z_{t} \sim N(0, \sigma_{z}^2)$ for $z \in \{ \varepsilon, \eta \}$. However, this fails to capture structural breaks in highly inflationary periods. To remedy the variability of the forecast \cite{stock2007} propose an additional dynamic to the log variance (also called volatility) as follows:
\begin{align}
    \log \sigma_{\eta, t+1} = \log \sigma_{\eta, t} + \nu_{\eta, t} \\
    \log \sigma_{\varepsilon, t+1} = \log \sigma_{\varepsilon, t} + \nu_{\varepsilon, t}
\end{align}
where $\nu_{z,t} \sim N(0, \gamma)$ for $z \in \{ \varepsilon, \eta \}$. In \cite{stock2007} the only model parameter $\gamma$ is set a priori to $0.2$, but can also be estimated using techniques from \cite{andrieu2010particle} or \cite{chopin2013smc2}.

\subsubsection{Outlier adjustments} With respect to structural changes, this model can accurately identify transitory effects with relative grace. However, large enough outliers incorrectly imply larger volatility spikes, when the measurement error is not necessarily a result of structural change. To account for these one-time measurement shocks, \cite{stock2016} suggest an alteration in the observation equation, where $\eta_{t} \sim N(0, s_{t}\sigma_{\eta, t}^2)$ such that $s_{t} \sim U(0,2)$ with probability $p$, or $s_{t} = 1$ otherwise.

\subsection{Integration with SSMProblems.jl}

Both UCSV (local level trend with stochastic volatility) and UCSV-O (UCSV with outlier adjustments) are similar models in nature, which facilitates an efficient construction of both models. Beginning with the common components, we define a class of latent dynamics which encompass the identical characteristics such as the transition log density.
\myjuliainput[][firstline=1,lastline=9]{code/TRENDINFLATION.jl}{Local level trend dynamics}{Local level trend dynamics}


An intuitive approach is to define model specific objects that dispatch differently for the transition dynamics. One can take a step further and define routines for local level trend dynamics and stochastic volatility dispatched on the parent class \texttt{LocalLevelTrend}. For brevity, we define the transition dynamics only for UCSV-O, since UCSV differs only the removal of \texttt{state[4]}.
\myjuliainput[][firstline=11,lastline=25]{code/TRENDINFLATION.jl}{UCSV-O transition dynamics}{UCSV-O transition dynamics}


Alternatively, for UCSV, trend dynamics dispatch on the class \texttt{SimpleTrend} which contains only a single field for $\gamma$. The full definition of both model dynamics are included in the examples folder of GeneralisedFilters.jl.

For the observation processes, both models can be defined using the \texttt{SSMProblems.distributions} constructor in the following code block.
\myjuliainput[][firstline=27,lastline=41]{code/TRENDINFLATION.jl}{Observation process definitions}{Observation process definitions}




Lastly, once the user defines the transition and observation processes, model construction is trivial using the built in \texttt{StateSpaceModel} constructor. This flexible interface also allows the user to specify the element types of the state depending on the chosen parameters; which is essential for compatibility with automatic differentiation frameworks, as well as single precision filtering algorithms.
\myjuliainput[][firstline=43,lastline=53]{code/TRENDINFLATION.jl}{Model construction}{Model construction}


\subsection{Results}

Instead of the MCMC samplers used in both \cite{stock2007} and \cite{stock2016}, we aim to replicate the results using a particle filter, where smoothed estimates are extracted from the sparse ancestry storage of \cite{jacob2015path}. Data for this case study is publicly available on FRED, where we query quarterly PCE inflation at a compounded annual rate of change\footnote{The formula for this transformation is $100 * \left(\frac{x_{t}}{x_{t-4}}-1 \right)$}.

We use a bootstrap filter with $N=2^{14}$ particles, resampling at every iteration with a Systematic resampling scheme. \texttt{GeneralisedFilters.jl} houses all of the required algorithms and interfaces with \texttt{SSMProblems.jl}; thus requiring minimal work from the user.

\begin{figure}[htp!]
    \centering
    \includegraphics[width=0.8\linewidth]{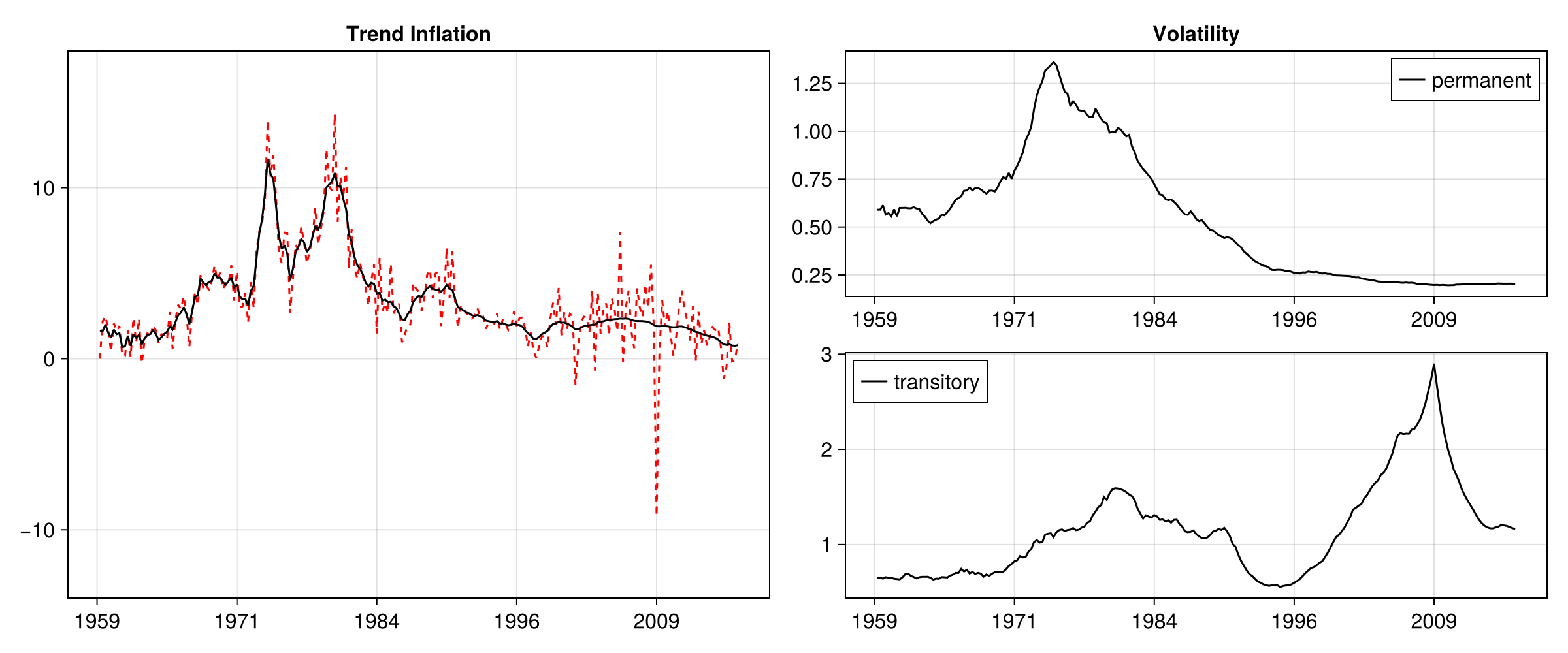}
    \caption{Unobserved components model with stochastic volatility (UCSV) estimates in black and observed data in red}
    \label{fig:UCSV result}
\end{figure}

Figure \ref{fig:UCSV result} depicts the approximately smoothed states from the UCSV model, and supplies visual evidence for the structural break in trend inflation, as is seen by the increased volatility throughout the 70s. There is further visual evidence for a massive spike in transitory volatility around the GFC, where outlier adjustments would otherwise correct.

\begin{figure}[htp!]
    \centering
    \includegraphics[width=0.8\linewidth]{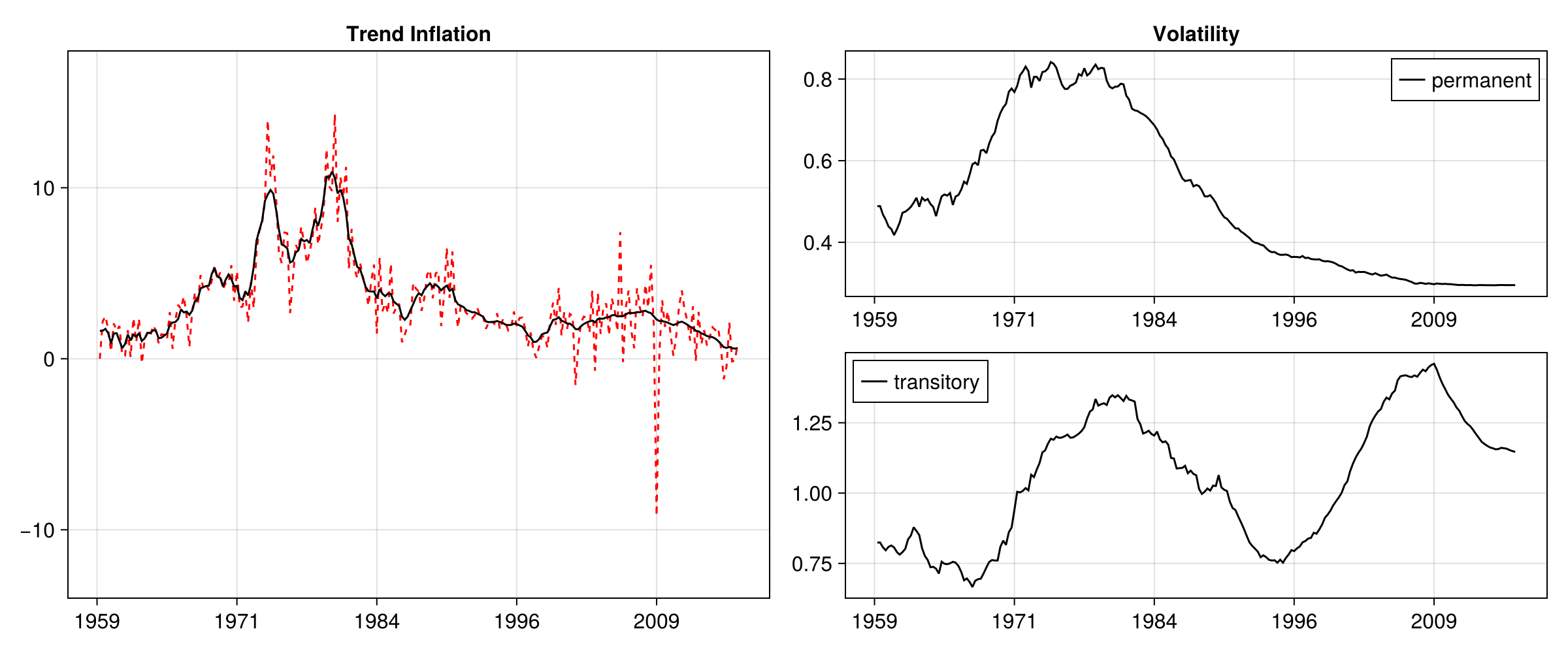}
    \caption{Unobserved components model with stochastic volatility and outlier adjustments (UCSV-O) estimates in black and observed data in red}
    \label{fig:UCSV-O result}
\end{figure}

With respect to UCSV-O, figure \ref{fig:UCSV-O result} provides a similar analysis with a dampened volatility spike in 2009. This model correctly identifies the structural break, without attributing all variability to measurement error. As a result, trend inflation exhibits more movement in the financial crisis, which \cite{stock2016} argue leads to better forecast errors.

\clearpage

\section{Supplementary materials for the data assimilation case study}

\subsection{Data Assimilation}
The goal of \textit{Data Assimilation} (DA) is to combine noisy and often incomplete observations of a system with a dynamical model to recover the true state of that system. DA has found numerous applications in the field of numerical weather modelling and geosciences amongst others, see \cite{carrassi2018data} for an introduction to these techniques. 

More formally, given a state trajectory $x_{1:T}$, noisy observations of the state of the form $y_{1:T} = H(x_{1:T}) + \eta_t $ and a model of the state evolution $p(x_{t+1} | x_t$) we try to recover the posterior $p(x_t | y_{1:t})$. In most applications the evolution model is deterministic and we need to introduce noise to the latent dynamics. 

\subsection{Lorenz 1963}
The Lorenz 63 model introduced in \cite{lorenz1963deterministic} is a simple model of atmospheric convection. The system describes the evolution of three variables $(x, y, z)$ which follow the non-linear system of ordinary differential equations:
\begin{align}
    &\dot{x} = \sigma(x - y) \\
    &\dot{y} = x(\rho - z) - y \\
    &\dot{z} = xy - \beta z
\end{align}
We set $\beta= 8/3$, $\rho=28$ and $\sigma=10$ for which the system is known to exhibit chaotic behaviour.

\subsection{Implementation in SSMProblems.jl}

We use the \texttt{ODEProblem} API in \texttt{OrdinaryDiffEq.jl} \cite{rackauckas2017differentialequations} to define the Lorenz system and integrate it numerically.
\myjuliainput[][firstline=1,lastline=20]{code/ASSIMILATION.jl}{DiffEq.jl problem definition and solver}{DiffEq.jl problem definition and solver}



The evolution model consists in one step of the numerical integrator above with added Gaussian noise
\[
    p(x_{t+1} | x_t) = \mathcal{N}(x_{t+1} | \mathcal{M}_t(x_t), \sigma^2)
\]
where $\mathcal{M}_t(x_t)$ is the solution of the system at time $t$. It is worth noting here that we have to reset all the internal states of the integrator as it will be shared across calls to \texttt{SSMProblems.distribution}.
\myjuliainput[][firstline=22,lastline=36]{code/ASSIMILATION.jl}{Solver iteration and noisy latent dynamics}{Solver iteration and noisy latent dynamics}




We observe the first component of the system and assume additive gaussian noise which leads to an observation process of the form
\[
    p(y_t | x_t) = \mathcal{N}(y_t | x^1_t, \nu^2)
\]
with $x^1_t$ the first component of the sytem at time $t$.
\myjuliainput[][firstline=38,lastline=44]{code/ASSIMILATION.jl}{Observation process}{Observation process}


Finally, both processes are used to define a \texttt{StateSpaceModel} 
\myjuliainput[][firstline=46,lastline=48]{code/ASSIMILATION.jl}{StateSpaceModel definition}{StateSpaceModel definition}

\subsection{Settings and results}
\begin{figure}[htp!]
    \centering
    \includegraphics[width=0.8\linewidth]{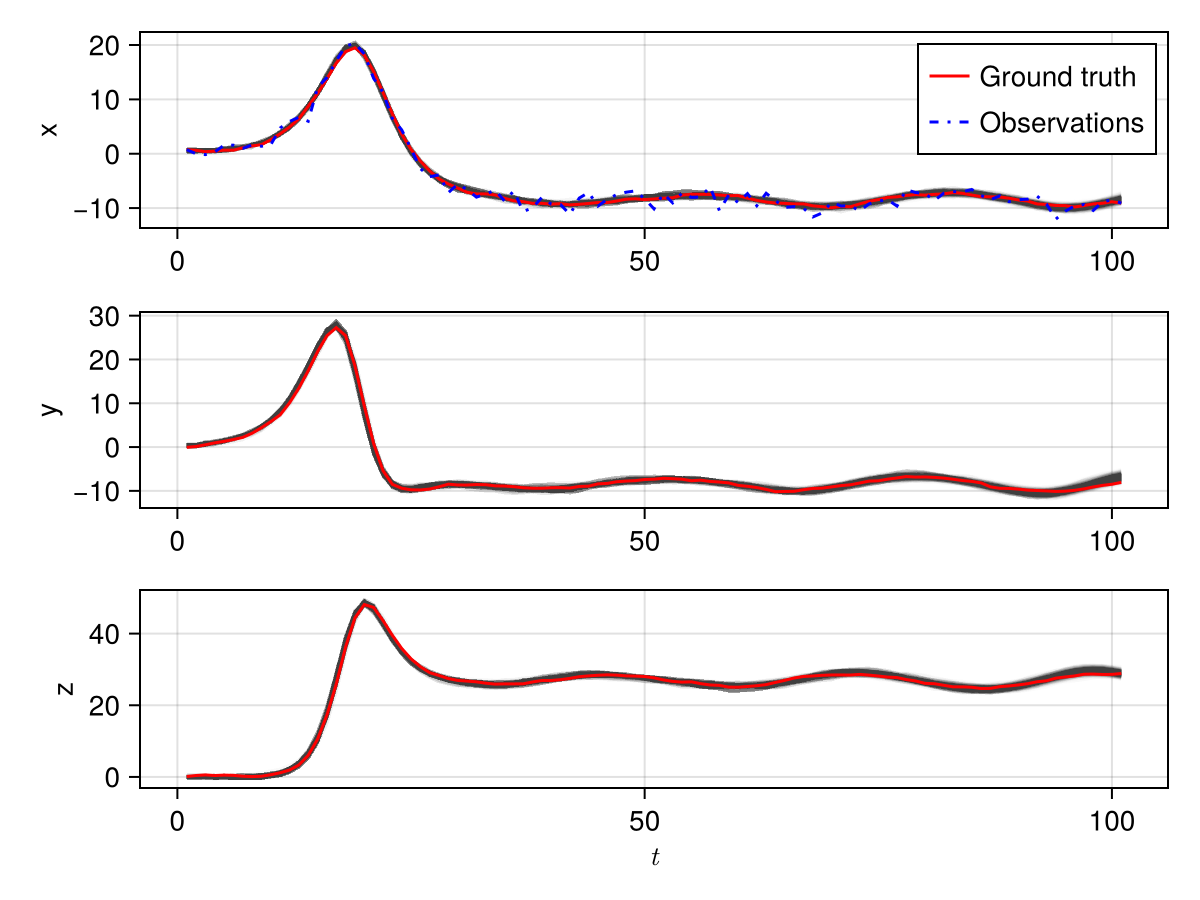}
    \caption{Unobserved components of the Lorenz 63 model with the estimated trajectories.}
    \label{fig:bf-da-results}
\end{figure}

We simulate a reference trajectory by solving the system for $N = 100$ steps of size $dt = 0.025$ with additive noise. A bootstrap particle filter implemented in \texttt{GeneralizedFilters.jl} with 1024 particles is used to estimate the latent state posterior. The ancestry paths of the particles  and the reference trajectories are shown on figure \ref{fig:bf-da-results}.

\clearpage


\end{document}